\documentclass[12pt]{article}
\usepackage{graphicx}

\newsavebox{\sboxpubnumber}
\newsavebox{\sboxpubdate}
\newcommand{\pubdate}[1]{\begin{lrbox}{\sboxpubdate}{#1}\end{lrbox}}
\newcommand{\pubnumber}[1]{\begin{lrbox}{\sboxpubnumber}{\begin{tabular}{l} #1 \\
				 \usebox{\sboxpubdate}
				 \end{tabular}}
                           \end{lrbox}
                           \pubblock}
\newcommand{\Title}[1]{\begin{center} {\Large #1 } \end{center}}
\newcommand{\Author}[1]{\begin{center}{ \sc #1} \end{center}}
\newcommand{\Address}[1]{\begin{center}{ \it #1} \end{center}}

\newcommand{\pubblock}{\rightline{
			\usebox{\sboxpubnumber}}}
\newenvironment{Abstract}{\begin{quotation}  }{\end{quotation}}
\newenvironment{Presented}{\begin{quotation} \begin{center}
             PRESENTED AT\end{center}\bigskip
      \begin{center}\begin{large}}{\end{large}\end{center}
      \end{quotation}}

\begin{document}

\begin{titlepage}
\pubdate{\today}                    
\pubnumber{IFIC/01-63} 

\vfill
\Title{Models of Quintessential Inflation}
\vfill
\Author{Konstantinos Dimopoulos\footnote{work done in collaboration with 
J.W.F.~Valle and supported by the Spanish grant PB98-0693 and by the 
European Commission RTN network HPRN-CT-2000-00148. The author is in 
Physics Department, Lancaster University, Lancaster LA1 4YB, U.K. 
since 01 October 2001.}}
\Address{
Astroparticle \& High Energy Physics Group\\
Instituto de Fisica Corpuscular (IFIC)\\
Universitat de Valencia/CSIC\\
Edificio Institutos de Paterna\\
Apartado de Correos 22085\\
46071 Valencia, Spain}
\vfill
\begin{Abstract}
A unified approach to quintessence and inflation is investigated with the use 
of a single scalar field. It is argued that successful potentials have to 
approximate a combination of exponential and inverse power-law decline in 
the limit of large values of the scalar field. A class of such potentials is 
studied analytically and it is found that quintessential inflation is indeed 
possible. Successful models, not involving more than two natural mass scales,
are obtained, which do not require fine-tuning of initial conditions and do
not result in eternal acceleration.
\end{Abstract}
\vfill
\begin{Presented}
    COSMO-01 \\
    Rovaniemi, Finland, \\
    August 29 -- September 4, 2001
\end{Presented}
\vfill
\end{titlepage}
\def\thefootnote{\fnsymbol{footnote}}
\setcounter{footnote}{0}

\newcommand{\lsim}{\mbox{\raisebox{-.9ex}{~$\stackrel{\mbox{$<$}}{\sim}$~}}}
\newcommand{\gsim}{\mbox{\raisebox{-.9ex}{~$\stackrel{\mbox{$>$}}{\sim}$~}}}


\section{Introduction}
Recent observations suggest that the Universe at present is dominated by
Dark Energy \cite{dark} and undergoes accelerated expansion \cite{accel}. 
This can be attributed to the presence of a non-vanishing cosmological 
constant $\Lambda$, which, however, has to be extremely fine-tuned 
\mbox{$\Lambda^2\sim 10^{-120}M_P^4$}, where 
\mbox{$M_P=1.22\times 10^{19}$GeV} is the Planck mass. Therefore, 
the existence of a cosmological
constant is not favoured by theorists, who prefer to account for the presence 
of Dark Energy by different means. One novel idea is to consider a dynamical,
time-varying \mbox{$\Lambda=\Lambda(t)$} \cite{Peebles:1988ek}. The most 
straightforward realization of this idea is to suppose that $\Lambda(t)$ is 
due to the potential energy of a scalar field, called quintessence 
\cite{quint}, which comes to dominate the energy density of the Universe at 
present and drive the latter into a late-time accelerated expansion. However, 
quintessence too suffers from fine-tuning problems \cite{Kolda:1999wq}, 
so it is not clear that it is a better alternative to the 
cosmological~constant. 

A compelling way to minimize the fine-tunings of quintessence is to link it 
with inflation. This seems natural since both quintessence and inflation are 
based on the same idea; namely that the Universe undergoes accelerated 
expansion when dominated by the potential energy density of a scalar field, 
which slowly rolls-down its almost flat potential. Unification of inflation 
and quintessence is achieved by using a single scalar field $\phi$ to drive 
both. There are many merits to quintessential inflation. Firstly, one avoids 
the introduction of yet again another unobserved scalar field, whose nature 
and origin are unaccounted for. Furthermore, a single theoretical framework
may be used to construct the scalar potential $V(\phi)$. Moreover, it is
possible to minimize the fine-tunings of quintessence by linking them with 
the ones of inflation and by introducing only few mass scales and parameters 
in $V(\phi)$. Finally, certain fine-tunings are automatically dispensed with,
such as, for example, the tuning of initial conditions for quintessence.

Apart from satisfying the requirements of inflation and quintessence,
quintessential inflation needs to incorporate an additional number of features.
One such requirement is that the scalar field should not be coupled to any of 
the standard model fields. This is so because it is necessary for $\phi$ to 
avoid decay and survive until today. The additional advantage of such a 
``sterile'' inflaton is that one dispenses with the fine-tuning of the 
couplings between $\phi$ and the standard model fields, otherwise unavoidable 
in order to preserve the flatness of the inflationary potential and also 
because the ultra-light quintessence field would correspond to a long-range 
force that could violate the equivalence principle at present. Natural 
candidates for a sterile inflaton are hidden sector fields, moduli fields, the 
radion and so on. Another requirement of quintessential inflation is that the 
minimum of the potential (taken to be zero, i.e. there is no residual 
cosmological constant) should not have been reached until today, so that 
$V(\phi)$ may dominate the Universe at present. In order to achieve this, the 
minimum is typically placed at infinity. Thus, the potential features two flat 
regions, the inflationary plateau and the quintessential ``tail''.

It is not easy to construct successful quintessential inflationary models
and this is why only few such models exist in the literature. Most of the
existing models, however, manage to satisfy the requirements of inflation
and quintessence by considering multi-branch potentials that change form
when the field moves from the inflationary to the quintessential phase of its 
evolution. This is achieved either ``by hand'' \cite{q-inf-1} or by a phase 
transition \cite{q-inf-2}. In these models information is not communicated 
between the inflationary and the quintessential part of the field's evolution. 
Another option, is constructing complicated models involving a large number of 
parameters, usually due to the theoretical framework used, which have to be 
tuned correctly to achieve the desired results \cite{q-inf-3}. 

We adopt a different approach and attempt to formulate a single-branch 
potential with minimal parameter content \cite{Dimopoulos:2001ix}. Such a 
minimalistic approach avoids the large number of fine-tunings inherent in 
other models, and renders our quintessential inflationary models preferable 
compared to the cosmological constant alternative.
\section{Requirements of inflation and quintessence}
Recent CMB observations \cite{cmbobs} suggest that we live in a 
spatially-flat FRW Universe. We model the Universe content as a collection of 
perfect fluids, the background fluid with density $\rho_B$ comprised by matter 
(including baryons and CDM) and radiation (including relativistic particles) 
and the scalar field $\phi$ with density 
\mbox{$\rho_\phi\equiv\rho_{\rm kin}+V$} and pressure 
\mbox{$p_\phi\equiv\rho_{\rm kin}-V$}, where 
\mbox{$\rho_{\rm kin}\equiv\frac{1}{2}\dot{\phi}^2$}. The evolution of $\phi$ 
is determined by the equation,
\begin{equation}
\ddot{\phi}+3H\dot{\phi}+V'=0
\label{feq}
\end{equation}
where \mbox{$H\equiv\dot{a}/a$} is the Hubble parameter with $a(t)$ being the
scale factor of the Universe and the prime \{dot\} denotes derivative with 
respect to $\phi$ \{the cosmic time $t$\}. 
\subsection{Inflationary requirements}
During inflation $V(\phi)$ dominates the Universe and the above equation 
becomes,
\begin{equation}
3H\dot{\phi}\simeq -V'(\phi)
\label{field}
\end{equation}
Inflation occurs when $|\varepsilon|,|\eta|<1$, where,
%
\mbox{$\varepsilon\equiv\frac{1}{\sqrt{6}}\,m_P(V'/V)$}
and \mbox{$\eta\equiv\frac{1}{3}\,m_P^2(V''/V)$}.
%
\subsubsection{Horizon and flatness problems}
The horizon and flatness problems are solved if the scale that corresponds to 
the observable Universe at present did exit the horizon during inflation. The 
number of e-foldings before the end of inflation when this happened is 
estimated as \mbox{$N_H\simeq\ln(t_0T_{_{\rm \!CMB}})-N_{\rm reh}$},
where $t_0$ is the age of the Universe according to the Hot Big Bang, 
$T_{_{\rm \!CMB}}$ is the temperature of the CMB at present and 
\mbox{$N_{\rm reh}\equiv\ln(T_{\rm reh}/H_{\rm end})$} with $T_{\rm reh}$
being the reheating temperature and $H_{\rm end}$ the Hubble parameter at 
the time $t_{\rm end}$ when inflation ends.

To solve the horizon and flatness problems the total number of inflationary 
e-foldings has to be larger than $N_H$, which translates into a constraint 
on the initial value of $\phi$.
\subsubsection{COBE normalization}
This constraint corresponds to the amplitude of the density perturbations
as deduced by the CMB temperature anisotropies observed by COBE. The constraint
reeds,
\begin{equation}
\left.\frac{\Delta T}{T}\right|_{\rm dec}\simeq
\frac{\delta\rho}{\rho}(N_{_{\rm\!CMB}})\simeq 2\times 10^{-5}
\label{cobe}
\end{equation}
where 
\begin{equation}
\frac{\delta\rho}{\rho}\;\simeq\;
\left.\frac{H^2}{\pi\dot{\phi}}\,\right|_{\rm exit}
\simeq -\frac{1}{\sqrt{3}\pi}\frac{V^{3/2}}{m^3_PV'}
\label{dr/r}
\end{equation}
and \mbox{$N_{_{\rm\!CMB}}\simeq\ln(t_{\rm dec}T_{\rm dec})-N_{\rm reh}$},
with \mbox{$T_{\rm dec}\equiv T(t_{\rm dec})$} being the temperature of the
last scattering surface and $t_{\rm dec}$ being the time of decoupling between
matter and radiation. Typically, the COBE normalization constraint determines 
the inflationary energy scale.
\subsubsection{Spectral index}
It can be shown that the spectral index $n_s$ of the density perturbations is
given by,
\begin{equation}
n_s-1\simeq 6(\eta-3\varepsilon^2)
\label{n}
\end{equation}
Large scale structure and 
CMB observations provide the following constraint for $n_s$,
\begin{equation} 
|n(N_{_{\rm\!CMB}})-1|\leq 0.1
\end{equation}
which constraints the slope and curvature of the potential $|V'|$ 
and $|V''|$ respectively.
\subsection{Quintessential requirements}
\subsubsection{Coincidence}
The scalar field should account for the required Dark Energy at present.
Thus, $\rho_\phi$ needs to be comparable with $\rho_B$ today, but subdominant 
during the Hot Big Bang. The constraint reeds, 
\mbox{$V(\phi_0)=\Omega_\phi\rho_0$}, where \mbox{$\phi_0\equiv\phi(t_0)$}, 
$\rho_0$ is the overall energy density of the Universe at present and 
\mbox{$\Omega_\phi\approx 2/3$} is the currently observed abundance of Dark 
Energy \cite{dark}. 
\subsubsection{Acceleration}
The SN Ia observations suggest that the Universe at present is 
undergoing accelerated expansion \cite{accel}. In order to achieve this we 
need \mbox{$\rho_\phi(t_0)>\rho_B(t_0)$} and also 
\mbox{$-1\leq w_\phi(t_0)<-\frac{1}{3}$}, where 
\mbox{$w_\phi\equiv p_\phi/\rho_\phi$}. 
String theory considerations disfavor eternal acceleration because it results
in the existence of a future event horizons \cite{strings}. If we take this 
constraint into account we can allow only for a brief acceleration period 
occurring today.
\section{Evolution}
The particular characteristics of quintessential inflationary potentials 
result in a particular scenario for the Universe evolution, which imposes a
few additional constraints.
\subsection{After the end of inflation}
Reheating, in quintessential inflation, is due to the gravitational production 
of non-conformally invariant, effectively massless fields, facilitated by the 
change of the vacuum at the end of inflation \cite{Ford:1987sy}. The reheating 
temperature is, \mbox{$T_{\rm reh}=\alpha(H_{\rm end}/2\pi)$}, where 
\mbox{$\alpha\sim 0.01$} is an (in)efficiency factor \cite{Joyce:1998fc}. 
Therefore, \mbox{$N_{\rm reh}=\ln(\alpha/2\pi)\simeq -6$} regardless of the 
inflationary energy scale, so that, \mbox{$N_H\simeq 74$} and 
\mbox{$N_{_{\!\rm CMB}}\simeq 69$}.
\subsubsection{Kination}
Because, at the end of inflation, 
\mbox{$\rho_B(t_{\rm end})\sim T_{\rm reh}^4\ll V_{\rm end}$} the Universe 
remains $\phi$--dominated. However, after the end of slow roll we have 
kination with \mbox{$\rho_{\rm kin}\gg V$} 
\cite{Joyce:1998fc} and (\ref{feq}) becomes,
\begin{equation}
\ddot{\phi}+(\dot{\phi}/t)\simeq 0
\label{kinfield}
\end{equation}
Thus, the evolution of $\phi$ is independent of $V(\phi)$. 
Because \mbox{$\rho_{\rm kin}\propto a^{-6}$} soon $\rho_B$ comes
to dominate and the Hot Big Bang begins. The temperature at the end of 
kination is,
\begin{equation}
T_*=\pi\sqrt{\frac{g_*}{30}}\,\frac{T_{\rm reh}^3}{V_{\rm end}^{1/2}}=
\frac{\alpha^3}{72\pi^2}\sqrt{\frac{g_*}{30}}\,\frac{V_{\rm end}}{m_P^3}
\label{T*}
\end{equation}
where \mbox{$V_{\rm end}\equiv V(t_{\rm end})$} and $g_*$ is the number of 
relativistic degrees of freedom, which, for the standard model in the early 
Universe, is \mbox{$g_*=106.75$}. Kination has to be over before Big Bang 
Nucleosynthesis (BBN). Therefore, \mbox{$T_{_{\!\rm BBN}}<T_*$}, where 
\mbox{$T_{_{\!\rm BBN}}\simeq 0.5$ MeV} is the temperature at the onset of BBN.
The BBN constraint provides a lower bound on $V_{\rm end}$.
\subsubsection{Hot Big Bang}
After the end of kination \mbox{$\rho_{\rm kin}\propto t^{-3}$} so that
\mbox{$\dot{\phi}\rightarrow 0$} rapidly. Consequently, the scalar field
freezes at some value $\phi_F$, estimated as,
\begin{equation}
\phi_F\simeq\phi_{\rm end}+2
\mbox{\Large $\sqrt{\frac{2}{3}}$}
\left[1+\frac{3}{2}\ln\Big(\frac{12\pi}{\alpha^2}
\mbox{\Large $\sqrt{\frac{30}{g_*}}$}
\,\Big)+3\ln\Big(\frac{m_P}{V_{\rm end}^{1/4}}\Big)\right]\,m_P\gg m_P
\label{fF}
\end{equation}
where \mbox{$m_P\equiv M_P/8\pi$} and 
\mbox{$\phi_{\rm end}\equiv\phi(t_{\rm end})$}. 
\subsection{Attractors and trackers}
During the quintessential part of its evolution the scalar field is again
dominated by its potential energy density so that (\ref{field}) is still
applicable. However, during the Hot Big Bang, \mbox{$H\propto t^{-1}$}
and the roll of the field is not as much restrained as in inflation.
\subsubsection{Attractor solution}
The solution of (\ref{field}) is of the form,
\mbox{$f(\phi\mbox{\small $(t)$})-f(\phi_F)=F(t)$},
where \mbox{$F(t)\equiv\frac{1}{4}(1+w_B)t^2$} and 
\mbox{$f(\phi)\equiv-\int'd\phi/V'$} and the prime on the integral means that 
one should not consider constants of integration. The above suggests
\begin{equation}
\phi(t)\simeq\left\{
\begin{array}{lrl}
\phi_F & \mbox{when}\hspace{0.5cm}
f(\phi_F)\gg F(t) & \;\mbox{\ Frozen Quintessence}\\
 & & \\
\phi_{\rm atr}(t) & \mbox{when}\hspace{0.5cm}
f(\phi_F)\ll F(t) & \;\mbox{\ Attractor Quintessence}
\end{array}
\right.
\end{equation}
where $\phi_{\rm atr}(t)$ is the solution of 
\begin{equation} 
f(\phi)\simeq F(t)\Leftrightarrow 
\frac{1}{4}(1+w_B)t^2 =-\int'\frac{d\phi}{V'(\phi)}
\label{atr}
\end{equation}
referred to as attractor solution \cite{track}. Because $F(t)$ is a 
growing function of time, even though initially the field remains frozen at 
$\phi_F$, later on it unfreezes and starts following the attractor. This 
occurs when 
\mbox{$V_{\rm atr}\equiv V(\phi_{\rm atr})\simeq V_F\equiv V(\phi_F)$} 
so that, at all times,
\mbox{$V(\phi)=$min\{$V_F,V_{\rm atr}$\}}. Note that the 
attractor solution is independent of initial conditions.
\subsubsection{The choice of the quintessential tail}
Depending on the slope of the quintessential tail $V_{\rm atr}$ may fall more 
\{less\} rapidly than $\rho_B$. We will call such an attractor steep \{mild\}. 
If the attractor is steep, only frozen-quintessence may achieve coincidence. 
However, steep attractors begin soon after the end of inflation. Hence, 
potentials with steep quintessential tails are ruled out. Mild attractors are 
different. Such attractors assist $\phi$ to eventually dominate $\rho_B$ and, 
hence, they are called ``trackers'' \cite{track}. Mild quintessential tails 
and all cases of frozen-quintessence result in eternal acceleration, 
disfavored by string theory \cite{strings}, and are, therefore, also ruled 
out.\footnote{Quintessential tails, which change slope from mild to steep 
\cite{hybridtail}, may be constructed only at the expense of additional 
mass-scales and parameters, which contrasts our minimalistic approach.} 

\begin{figure}[htb]
    \centering
    \includegraphics[height=2.5in]{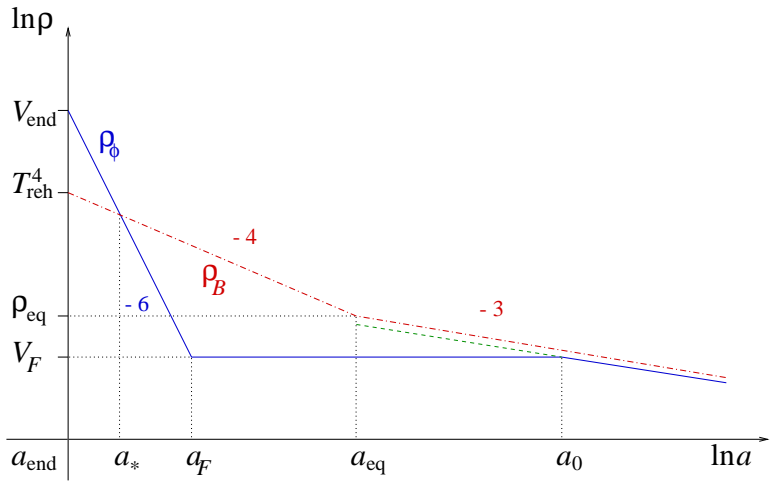}
\caption{Exponential attractor (dashed line). After the end of kination,
$\rho_\phi$ (solid line) remains frozen until today, when it begins to mimic 
the evolution of $\rho_B$ (dash-dot line).}
    \label{q2}
\end{figure}
\begin{figure}[htb]
    \centering
    \includegraphics[height=3.5in]{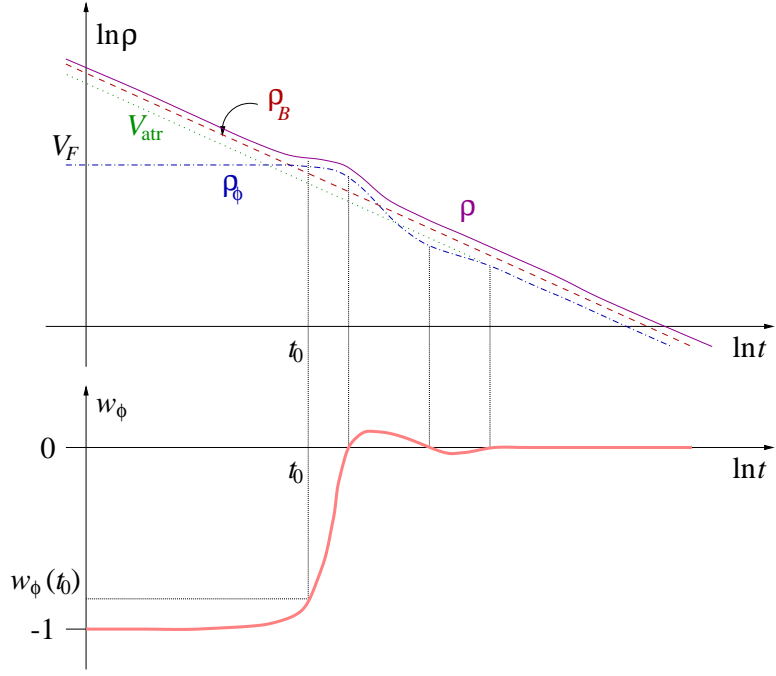}
    \caption{At unfreezing, $\rho_\phi$ (dash-dot line) briefly oscillates
around the attractor (dotted line), which mimics $\rho_B$ (dashed line). 
Consequently, there is a ``bump'' on the overall density $\rho$ (solid line).
During the ``bump'', \mbox{$w_\phi\simeq -1$}, which results into brief 
acceleration.}
    \label{q4}
\end{figure}

Therefore, the only acceptable case is to consider attractors according to 
which $V_{\rm atr}$ falls as rapidly as $\rho_B$ (see Fig. \ref{q2}), so that 
\mbox{$V_{\rm atr}/\rho_B=$ constant.} This corresponds to exponential 
quintessential tails \cite{exp-q}, for which the attractor (\ref{atr}) in the 
matter era is,
\begin{equation}
\frac{\rho_B}{V_{\rm atr}}=4\varepsilon^2={\rm constant}
\label{eps3}
\end{equation}
Although it may explain the missing Dark Energy, the exponential attractor
is seemingly unable to result in accelerated expansion, because it mimics 
$\rho_B$ and, hence, \mbox{$w_\phi(t_0)\simeq 0$}. However, a brief period of 
acceleration is indeed possible to achieve if the field unfreezes at present 
\cite{exp-new}. This is because, when unfreezing, the field oscillates briefly 
around the attractor. Thus, at first crossing of $V_F$ and $V_{\rm atr}$ the 
field remains ``super-frozen'' and dominates the Universe, causing accelerated 
expansion. Soon, however, it settles onto the attractor and acceleration is 
terminated (Fig. \ref{q4}). Thus, coincidence requires 
\mbox{$\phi_{\rm atr}(t_0)\simeq\phi_F$}. Moreover, ``super-freezing'' results 
in \mbox{$w_\phi(t_0)\simeq -1$} which guarantees acceleration.

Therefore, we are led to try a quintessential tail of the form,
\mbox{$V(\phi\gg\phi_{\rm end})\simeq V_{\rm end}\,e^{-\lambda\phi/m_{\!P}}$}
where \mbox{$\lambda>0$} is a parameter for which
\mbox{$\varepsilon=-\lambda/\sqrt{6}$}. In order to avoid frozen quintessence
we require \mbox{$V_{\rm atr}\lsim\rho_B$}, which means that $\lambda$ should 
be close but not smaller than {\small $\sqrt{3/2}$}. However, it turns out 
that the corresponding value for $V_{\rm end}$ cannot satisfy the BBN 
constraint. Fortunately, one can overcome this difficulty by modifying the 
the quintessential tail in a way that preserves the exponential attractor. 
This is achieved by introducing an Inverse Power-Law (IPL) factor so that,
\mbox{$V(\phi\gg\phi_{\rm end})\simeq 
V_{\rm end}\,e^{-\lambda\phi/m_{\!P}}(m/\phi)^k$}, where \mbox{$k\geq 1$} is 
an integer and \mbox{$m\leq m_P$}. The attractor for this quasi-exponential 
tail is identical with the pure exponential case because 
\mbox{$\phi_F\gg m_P$} suggests,
\begin{equation}
V'(\phi_F)=-\frac{V}{m_P}\Big[\lambda+k\Big(\frac{m_P}{\phi_F}\Big)\Big]
\simeq-\frac{\lambda V}{m_P}
\end{equation}
The quasi-exponential behaviour cannot carry over to the inflationary era 
because of the steepness it results to. Thus, both the exponential and the IPL 
features have to be modified in inflation. This modification is not trivial 
because of the huge difference \mbox{$V_{\rm end}\gg V_F\sim\rho_0$} as 
required by BBN and coincidence. A steep inflationary plateau results either 
in too brief inflation or in strongly super-Planckian inflationary energy 
scale. A flat inflationary plateau, however, because it has to ``prepare'' for 
the deep dive towards $V_F$, typically features a large value of $|V''|$, 
which results in too large $n_s$ [c.f.~(\ref{n})].
\section{A concrete example}
\begin{figure}[htb]
    \centering
    \includegraphics[height=2.25in]{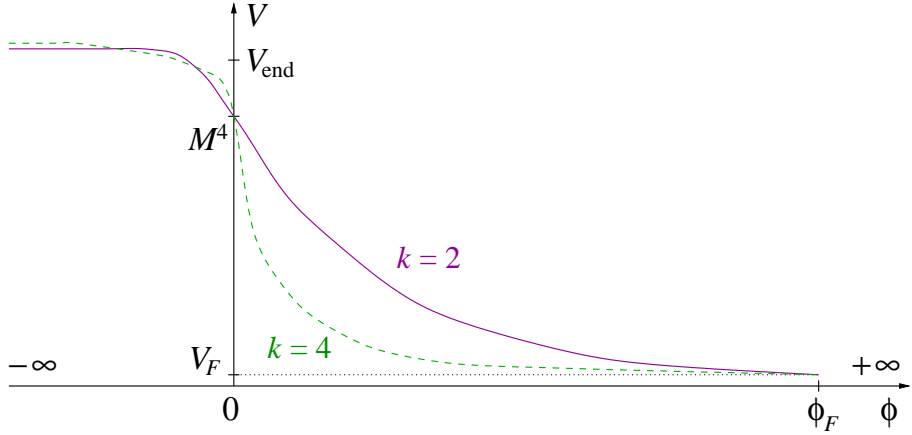}
\caption{Artist's view of model (\ref{model}) for $k=2,4$ (solid,dashed line
respectively).}
    \label{q3}
\end{figure}

Designing a quintessential inflationary potential is not easy, let alone using 
few mass scales and parameters. Nevertheless, it is indeed possible. For 
example, consider the potential,

\begin{equation}
V(\phi)=M^4[1-\tanh(\phi/m_P)]
\Big[1-\sin\Big(\frac{\pi\phi/2}{\sqrt{\phi^2+m^2}}\Big)\Big]^k
\label{model}
\end{equation}
where \mbox{$-\infty<\phi<+\infty$}, \mbox{$M,\:m<m_P$}  and \mbox{$k>0$} is 
an integer (Fig.~\ref{q3}). The asymptotic forms of the above far from the 
origin are, 
\begin{equation}
V(\phi\ll 0)\simeq 2^{k+1}M^4
\left[1-e^{2\phi/m_P}-\frac{k\pi^2}{64}(\frac{m}{\phi})^4\right]
\simeq 2^{k+1}M^4
\end{equation}
and
\begin{equation}
V(\phi\gg 0)\simeq 2^{1-k}(\frac{\pi}{4})^{2k}e^{-2\phi/m_P}M^4
\Big(\frac{m}{\phi}\Big)^{4k}\propto
\,e^{-2\phi/m_{\!P}}(m/\phi)^{4k}
\end{equation}
Thus, for negative values of $\phi$ the potential approaches a 
constant, non-zero, false vacuum energy density responsible for inflation, 
whereas for positive values of $\phi$ the potential attains the
desired quasi-exponential form with \mbox{$\lambda = 2$}.
Enforcing the constraint (\ref{cobe}) gives,
\begin{equation}
\frac{2^{\frac{k+1}{2}}}{\sqrt{3}\,\pi}\Big(\frac{M}{m_P}\Big)^2
(N_{_{\!\rm CMB}}+1/3)=10^{-5}\Rightarrow
 M\sim 10^{15}\mbox{GeV}
\end{equation}
Thus, we see that $M$ is of the scale of grand unification. Now, employing
(\ref{n}) we obtain,
\begin{eqnarray}
n_s(N)-1\simeq-\frac{6}{3N+1}\left[1+\frac{9}{8(3N+1)}\right]
\Rightarrow n_s(N_{_{\!\rm CMB}})\simeq 0.97
\nonumber
\end{eqnarray}
which is in excellent agreement with the observations \cite{cmbobs}. 
Also, \mbox{$V_{\rm end}\simeq 2^{k-1}M^4$} gives,
\begin{equation}
T_{\rm reh}\simeq\frac{\alpha}{4}\frac{10^{-5}m_P}{\sqrt{N_{_{\!\rm CMB}}+1/3}}
\sim 10^{9}\,\mbox{GeV}
\end{equation}
which saturates but does not violate the gravitino constraint. Using the above,
(\ref{T*}) gives,
\begin{equation}
T_*\simeq\frac{\alpha^3\pi^2}{96}\sqrt{\frac{g_*}{30}}
\frac{10^{-10}m_P}{(N_{_{\!\rm CMB}}+1/3)^2}\sim 10\;\mbox{MeV}> 
T_{_{\!\rm BBN}}
\end{equation}
Thus, the BBN constraint is satisfied. Finally, solving the horizon and 
flatness problems requires the initial condition for the inflaton,
\mbox{$\phi_i\leq\phi(N_H)\simeq -3\,m_P$}, which does not require fine-tuning
since we expect \mbox{$\phi_i\sim|M_P|$}. Note that, $n_s,T_{\rm reh},T_*$
are all $k$-independent.

Moving on to quintessential requirements, from (\ref{atr}), the attractor for 
the matter era is, \mbox{$V_{\rm atr}/\rho_B=3/8$}.
Super--freezing is expected to boost this fraction up to 
\mbox{$\Omega_\phi\approx 2/3$} and also to give \mbox{$w_\phi\approx -1$}.
The freezing value of $\phi$ is found using (\ref{fF}),
\begin{equation}
\phi_F=\sqrt{6}\left\{\frac{2}{3}-
\frac{1}{\sqrt{6}}\ln(2/\sqrt{3})+
\ln\left[\frac{24}{\alpha^2}\mbox{\Large $\sqrt{\frac{10}{g_*}}$}
\Big(N_{_{\!\rm CMB}}+\frac{1}{3}\Big)\times 10^5\right]\right\}m_P\simeq 67m_P
\end{equation}
Using the coincidence constraint \mbox{$\phi_0\simeq\phi_F$} and after some 
algebra one finds,
\begin{equation}
m\sim 4\times 10^2(4\times 10^{70})^{1/4k}10^{-30/k}\,m_P
\end{equation}
which gives,
\begin{center}
\begin{tabular}{|c||c|c|c|c|}
\hline
$k$ & 1 & 2 & 3 & 4 \\ \hline
$m$ & $10^8$GeV & $10^{15}$GeV & $10^{17}$GeV & $10^{18}$GeV\\
\hline
\end{tabular}
\end{center}
Thus, we can identify $m$ with $M,m_P$ for \mbox{$k=2,4$} respectively, and 
obtain the models,
\begin{itemize}
\item
{\bf Model 1:}
\begin{equation}
V(\phi)=M^4[1-\tanh(\phi/m_P)]
\Big[1-\sin\Big(\frac{\pi\phi/2}{\sqrt{\phi^2+M^2}}\Big)\Big]^2
\end{equation}
\item
{\bf Model 2:}
\begin{eqnarray}\hspace{-1.5cm}
V(\phi)=M^4(\phi)\;[1-\tanh(\phi/m_P)]\hspace{0.2cm} & {\rm and} & 
\hspace{0.2cm}
M(\phi)\equiv M\Big[1-\sin\Big(\frac{\pi\phi/2}{\sqrt{\phi^2+m_P^2}}\Big)\Big]
\end{eqnarray}
\end{itemize}
\section{Conclusions}
We have shown that it is indeed possible to unify inflation and quintessence 
using a single scalar field without incorporating too many mass scales and 
parameters. The best approach suggests a quasi-exponential tail, which 
manages to meet BBN requirements, while retaining the pure-exponential 
attractor solution. The quasi-exponential tail achieves a brief acceleration 
period if the scalar field is about to unfreeze today and lies, at present, 
in a super-frozen state. Both the exponential and inverse-power law features 
of the tail have to be suppressed at the inflationary plateau.

Two successful models are presented, which satisfy all requirements of 
inflation and quintessence with the use of only two natural mass-scales: the
Planck and the grand unification scale, and no other parameters. Moreover, 
no fine tuning of initial conditions is required. In these models the 
inflationary scale is that of grand unification, which is low enough to 
ensure safety from radiative corrections. Also, 
\mbox{$T_{\rm reh}\sim 10^9$GeV}, which saturates the gravitino constraint 
and \mbox{$n_s\simeq 0.97$}, in excellent agreement with observations.
Finally, we expect \mbox{$w_\phi\approx -1$} because of super-freezing. 
Due to the plethora of constraints and requirements we believe that any 
successful model should not differ much from the toy-models presented.

In summary, successful quintessential inflationary models without additional 
fine-tuning for quintessence are possible to construct. Such models outshine 
the cosmological constant alternative.

\end{document}